\documentclass[10pt,twocolumn,twoside]{IEEEtran}
\IEEEoverridecommandlockouts
\usepackage{cite}
\usepackage{amsmath,amssymb,amsfonts}
\usepackage{graphicx}
\usepackage{textcomp}
\usepackage{xcolor}
\usepackage{comment}
\usepackage{algorithm}          
\usepackage{algpseudocode}

\usepackage{etoolbox}
\makeatletter
\patchcmd{\appendices}{\clearpage}{\par\vspace{-2ex}}{}{}
\makeatother

\algrenewcommand\algorithmiccomment[1]{\hfill // #1}
 
\usepackage{subcaption}
\usepackage{amsmath}
\def\BibTeX{{\rm B\kern-.05em{\sc i\kern-.025em b}\kern-.08em
    T\kern-.1667em\lower.7ex\hbox{E}\kern-.125emX}}
\begin{document}

\title{Predict, Reposition, and Allocate: A Greedy and Flow-Based Architecture for Sustainable Urban Food Delivery\\
{
}
\thanks{This work was supported by the Council of Scientific \& Industrial Research (CSIR) under Grant No. 22WS(0014)/2023-24/EMR-II/ASPIRE.}
}

\author{\IEEEauthorblockN{Aqsa Ashraf Makhdomi\IEEEauthorrefmark{1} and
Iqra Altaf Gillani\IEEEauthorrefmark{2}}
\IEEEauthorblockA{\\Department of Information Technology,
NIT Srinagar\\
Email: \IEEEauthorrefmark{1}makhdoomiaqsa@gmail.com,
\IEEEauthorrefmark{2}iqraaltaf@nitsri.ac.in}}

\maketitle

\begin{abstract}
The rapid proliferation of food delivery platforms has reshaped urban mobility but has also contributed significantly to environmental degradation through increased greenhouse gas emissions. Existing optimization mechanisms produce sub-optimal outcomes as they do not consider environmental sustainability their optimization objective. 
This study proposes a novel eco-friendly food delivery optimization framework that integrates demand prediction, delivery person routing, and order allocation to minimize environmental impact while maintaining service efficiency.
Since recommending routes 
is NP-Hard, the proposed approach utilizes the submodular and monotone properties of the objective function and designs an efficient greedy optimization algorithm. Thereafter, it formulates order allocation problem as a network flow optimization model, which, to the best of our knowledge, has not been explored in the context of food delivery. A three-layered network architecture is designed to match orders with delivery personnel based on capacity constraints and spatial demand. Through this framework, the proposed approach reduces the vehicle count, and creates a sustainable food delivery ecosystem. 

\end{abstract}

\begin{IEEEkeywords}
food delivery platforms, greedy algorithm, network flow, prediction, efficiency 
\end{IEEEkeywords}

\section{Introduction}  

Food delivery platforms have fundamentally transformed the dynamics of urban mobility systems by enabling on-demand access to meals, with just a few taps on their device.
Traditionally, individuals relied on dine-in or takeaway options, which restricted their choices to nearby restaurants and required additional time for travel and waiting. The emergence of digital platforms, combined with advancements in mobile applications, real-time tracking, and logistics, has disrupted this conventional model, making food delivery an integral component of modern urban systems. 
Despite these advancements, food delivery platforms pose significant environmental challenges. These platforms contribute substantially to greenhouse gas emissions due to the large number of delivery vehicles operating in urban areas. 
Studies indicate that vehicular emissions are a leading factor in declining air quality, particularly in high-density urban regions, where air pollution has been linked to a reduction in life expectancy by up to $10$ years \cite{Article:Airpoll_2019}. Given the increasing dependence on food delivery services, these platforms cannot be eliminated from society, but measures can be taken 
to design novel optimization mechanisms that balance operational efficiency with sustainability. 

This paper addresses this challenge by proposing an eco-friendly food delivery system that reduces the number of vehicles operating on the road while maintaining customer satisfaction and platform profitability. Although prior works \cite{Li:IEEEGlobecom_2018,Zhang:Springer_2022,Chen:IEEEITS_2024} have focused on the design of eco-friendly food delivery systems, they yield sub-optimal outcomes due to their inability to predict future demand. This limitation results in inefficient vehicle cruising, increased fuel consumption, and higher emissions. To address this gap, the proposed approach predicts the passenger demand using the graph neural network-based approach described in \cite{Bhat:Cods_25} and thereafter introduces two key approaches: (1) delivery person repositioning based on predicted demand pattern, and (2) order bundling to optimize vehicle utilization. \looseness=-1

\textit{Delivery person repositioning.}
The spatial-temporal distribution of food delivery demand is inherently dynamic, with peak demand varying across different locations and times of the day. If delivery persons operate reactively — waiting for orders at random locations or moving based on incomplete information — it leads to inefficient vehicle movement, increased fuel consumption, and prolonged service time. The proposed model overcomes this issue by predicting the demand for food orders and thereafter it recommends the routes to the delivery persons and brings them to the high-demand areas. 
Since the problem of route recommendation is NP-Hard \cite{Ashraf_ES:2024}, the proposed approach  utilizes the submodular and monotone properties of the objective function and employs a greedy algorithm to recommend routes effectively. 
Through the employment of prediction and routing mechanism, the proposed approach guides the delivery persons to high-demand areas, which ensures that they are optimally positioned when orders arrive. \looseness=-1

\textit{Order allocation.}
A major inefficiency in food delivery platforms arises from single-order assignments, where each delivery person picks up and delivers a single order before receiving a new assignment. This approach increases the total number of trips, leading to higher vehicle miles travelled and a negative impact on the environment. The proposed model addresses this issue by assigning multiple compatible food orders to a single delivery person through a novel \emph{multi-layer network flow-based approach}. In this framework, delivery persons and orders are represented as nodes in a flow network, and the edges capture constraints such as vehicle capacity and spatio-temporal demand distribution. Although flow-based approaches have been applied in domains such as logistics and resource allocation, no prior work has modelled food order allocation through a network flow framework.   Our proposed framework utilizes this formulation to jointly capture customer demand, platform capacity, and delivery constraints, enabling a more sustainable and efficient food delivery system. 
The contributions can be summarized as follows:

\begin{itemize}
    \item 
We propose a joint route recommendation system and order allocation mechanism for food delivery platforms that integrates passenger demand prediction and creates a sustainable platform. \looseness=-1 
    \item 
To overcome the complexity of the route recommendation, we utilize the submodular and monotone properties of the objective function which reduce the search space and provide greedy as a framework for solving the problem efficiently.

    \item To the best of our knowledge, this is the first work to frame order dispatching in food delivery as a \emph{flow} optimization problem which 
allocates customer orders based on the capacity and cost constraints of the delivery platform. 

    \item 
Extensive simulations on the  Meituan dataset demonstrates superior performance by the proposed model. \looseness=-1

\end{itemize}

\section{Related work}
\label{sec:rw}
The related work of the proposed approach is two-fold: 1) Order allocation mechanisms, and 2) Route recommendation systems. \looseness=-1

\subsection{Order allocation mechanisms}
Order allocation mechanisms play an important role in the efficient functioning of delivery platforms by assigning customer orders to the most suitable delivery personnel. These mechanisms are central to maximizing \textit{profit}, 
minimizing \textit{delivery time}, and creating \textit{ environmentally sustainable} system. Over the years, significant research has been conducted to design the order allocation mechanisms. 
This subsection provides a comprehensive overview of the advancements and methodologies developed in this domain. 

\textit{Profit maximization} is the primary objective of any operational platform, which ensures its long-term sustainability and competitive edge in the market. 
Considering the importance of this objective, it has received significant attention from the research community. In this direction, Li \textit{et al.} \cite{Li:IEEEGlobecom_2018} proposed
a non-cooperative sequential game theoretic framework wherein each worker was modelled as a player whose objective was to optimize his profit. However, 
their model primarily allocated tasks to high-speed, high-capacity delivery persons,  which led to the underutilization of other available workers and increased overall delivery cost. Moreover, their proposed approach did not account for customer order deadlines, which resulted in delay in servicing and reduced customer satisfaction with the system. 
Chen \textit{et al.}  \cite{Chen:IEEEITS_2024} proposed a graph neural network-based optimization algorithm for order dispatching in on-demand food delivery, that aimed to minimize delivery delay and improve system efficiency. It addressed the challenge of dynamically matching orders, riders, and restaurants in real-time under strict time constraints. Though this approach effectively captured spatio-temporal dependencies and outperformed traditional methods in large-scale scenarios, it required significant computational resources and assumed accurate input data, which may not always reflect real-world uncertainties like traffic or rider behavior. \looseness=-1

The second objective that has received significant attention is the \textit{delivery time} of the customer order. 
It represents the total time required to service an order, and is measured as the duration between order placement and its successful delivery at the customer's location. It should be minimized for ensuring the customer satisfaction with the system.
Various works have been done in the direction of reducing the customer delivery time. Lu \textit{et al.} \cite{Lu:ACMISMSI_2017} proposed an evolutionary algorithm for the order assignment in food delivery systems which reduced the distance travelled by delivery persons and resulted in the reduced customer delivery time. However, their method did not consider the allocation of multiple workers to orders originating from the same restaurant, which resulted in inefficiency in servicing.  Moreover, the computational complexity of their proposed model was high, which makes it infeasible for real-world practical scenarios. Ji \textit{et al.} \cite{Ji:WWW_2019} introduced a task grouping strategy
that accounted for order arrival times and assigned delivery persons to these groups to speed up the delivery process. However, their proposed approach ignored the spatial dimensions of order pickup and delivery locations, and did not incorporate real-time traffic conditions for estimating worker service times and associated costs. 

Apart from profit and expected delivery time, the third objective of platforms is the creation of an \textit{eco-friendly} system. 
In this direction, 
Liu \textit{ et al.} \cite{Liu:IEEETransMC_2019} proposed FooDNet, an optimized food delivery system that utilized spatial crowdsourcing and city taxis for dual passenger-food delivery. It uses an Adaptive Large Neighborhood Search algorithm to minimize the number of taxis, total distance, and delivery delay while maximizing delivery person income. Though this approach reduced cost and integrated efficiently with existing taxi operations, 
it was constrained by taxi availability and strict time windows for food delivery.
Liu \textit{ et al.} \cite{Liu:IEEETransITS_2024} proposed an eco-friendly on-demand meal delivery system using a mixed fleet of electric and gasoline vehicles, that was optimized through a rolling horizon framework with adaptive large neighborhood search. The proposed approach addressed the challenge of reducing greenhouse gas emissions while maintaining delivery efficiency by actively utilizing electric vehicles. 
However, their approach becomes computationally intensive for large-scale instances, limiting its practical applicability in dense environments. \looseness=-1 

The aforementioned works exhibit notable limitations that impact their overall efficiency and sustainability. First, these studies do not incorporate customer demand prediction, which results in the inefficient deployment of delivery persons to areas with low demand. This not only diminishes profitability but also leads to increased emissions, contributing to environmental degradation. Furthermore, these approaches do not simultaneously account for the delivery personnel's capacity and the balance between supply and demand during order allocation, which leads to suboptimal resource utilization.
In contrast, our proposed order allocation mechanism overcomes these challenges by introducing a multi-layered flow architecture. This framework integrates vehicular capacity, supply-demand equilibrium, and delivery person proximity when assigning customer orders to the delivery persons. Through this approach, our proposed model minimizes delivery time, maximizes profitability, and promotes an eco-friendly system. 

\subsection{Route recommendation}
Route recommendation systems are an important component of food delivery platforms, as these systems guide delivery persons by providing them the optimal path between order pickup and destination points. These systems aim to maximize \textit{profit}, 
reduce \textit{delivery time}, and create an \textit{eco-friendly} system. They are summarized next. 


\textit{Profit maximization} involves designing routing mechanisms that enhance the financial sustainability of the platform. 
In this direction,  
Chen \textit{et al.} \cite{Chen:IEEETransITS_2022} proposed an imitation learning-based iterated matching algorithm for optimizing on-demand food delivery process. It used a rolling horizon strategy to transform the dynamic problem into static generalized assignment problems and employed an offline-optimization for online-operation framework. 
They also proposed a route planning approach, which was treated as a pickup and delivery problem with a single vehicle and open route. Since the problem is NP-hard, they developed effective dispatching mechanism using the precomputed route costs and solved it in polynomial time.
 Huq \textit{et al.} \cite{Huq:IEEEAccess_2022} proposed a multi-objective linear programming framework that balanced worker profit and customer satisfaction. Since the proposed problem in NP-Hard, they proposed a metaheuristic solution based on water wave optimization, which enhanced service quality while providing additional worker incentives. 

\textit{Reducing delivery time } involves recommending routes to the delivery persons that will ensure quick servicing of customer orders.
In this direction,  Tu \textit{et al.} \cite{Tu:IEEEIOT_2020} formulated a mathematical model for the crowdsourced delivery problem, that aimed at minimizing total travel cost and delivery delay. Their proposed approach optimized rider-task assignments and delivery route selection by incorporating dynamic crowdsourcing with sequential order collection and rider allocation. However, due to the dynamic nature of order and rider arrival, the sequential assignment process can lead to suboptimal allocation, which causes significant delivery delay. Additionally, their model did not address long-term worker incentives, as it overlooked the profitability of delivery personnel. 
Furthermore, Chu \textit{et al.} \cite{Chu:SpringerCI_2023} proposed a data-driven framework  to address the last-mile problem in food delivery systems, which analyzed the worker behavior. However, this framework omitted the pickup phase of the ordered food, which limited its applicability in complex, real-world practical scenarios. \looseness=-1

\textit{Eco-friendly} route recommendation mechanisms direct routes to the delivery persons that reduce the greenhouse emissions released by the vehicles. 
In this direction, Liao \textit{et al.} \cite{Liao:Elsevier_2020} proposed a green meal delivery routing problem with multiple objectives, that focused on optimizing customer satisfaction and rider efficiency while reducing the carbon footprint. It addressed the problem of quantifying environmental impacts by integrating order assignment, fleet sizing, and routing optimization into a mixed-integer linear programming framework. While the approach provided valuable insights into emission reduction strategies and highlighted the benefits of electric vehicle adoption, it assumed simplified charging behavior which does not fully account for real-world complexities like varying traffic conditions or delivery person behavior. 
Caggiani 
\textit{et al.} \cite{Caggiani:IEEEConf_2020} proposed an eco-friendly route recommendation system for delivery persons operating on cargo bikes. It utilized a dynamic algorithm to compare routes based on minimum travel time and minimum emission exposure, and incorporated real-time traffic and emissions data. Their model improved delivery person well-being and sustainability, but it can slightly increase travel time and require complex real-time data integration.
Moreover, all of the above works recommend the routes to the delivery persons after the customer orders the food. In contrast, our proposed approach first predicts the customer demand using the graph neural network-based approach \cite{Bhat:Cods_25} and thereafter recommends the routes to the delivery persons based on the customer demand. 






\section{Proposed model}
\label{sec:pm}
This section describes the working of the proposed model. It first discusses the route recommendation system which will place the delivery persons in the areas where customer demand is expected to be high and thereafter follow up with the order allocation mechanism. 

\subsection{Route recommendation system}
This subsection begins with the graphical modelling approach used in the design of the route recommendation system, followed by the problem formulation and solution for the route recommendation system. 

\subsubsection{Graphical modelling}


The proposed route recommendation system is represented by a family of subgraphs, denoted as $\mathcal{G} = \{G^D, G^S\}$, where $G^D$ and $G^S$ represent distance and customer-order subgraphs, respectively. 
Each subgraph is defined as $G^i = (V^i, E^i,w^i),\, \, i\,\in\,\{D,S\}$, where $V^i$ represents the nodes in subgraph $i$, $E^i$ denotes the edges in subgraph $i$, and $w^i$ denotes the edge weight corresponding to the edge $E^i$. These subgraphs are detailed next.

\begin{figure}[t!]
\vspace*{-9mm}
    \centering
    \begin{subfigure}[b]{0.23\textwidth} \centering
        \includegraphics[width=0.395634\linewidth]{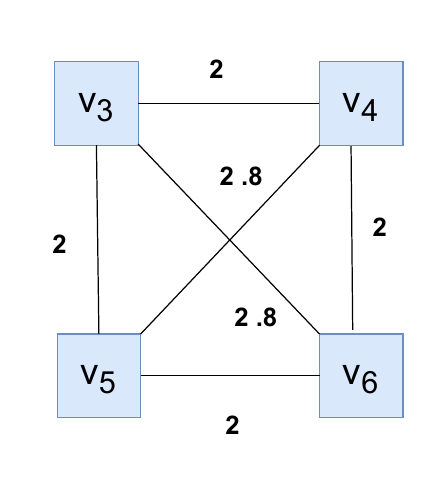}
        \vspace*{-2mm}
        \caption{}
        \label{img:road}
    \end{subfigure} \hfill
    \begin{subfigure}[b]{0.23\textwidth} \centering
        \includegraphics[width=0.395634\linewidth]{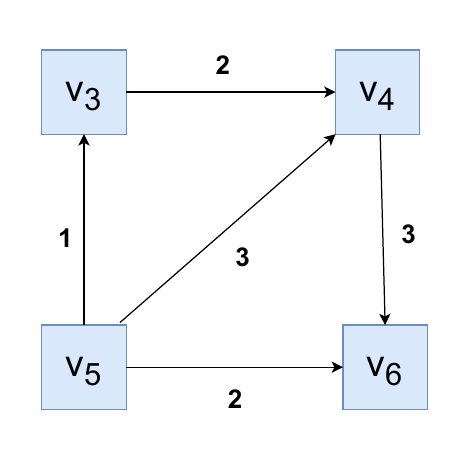}   \vspace*{-2mm}
        \caption{}
        \label{img:req}
    \end{subfigure}
    \vspace{-2mm}
   \caption{Subgraph family: (a) distance and (b) customer order}
    \label{utility}\vspace*{-5mm}
\end{figure}

\textit{Distance subgraph $G^D$.}
This subgraph models the structural layout of the road network.
Based on the directions from previous studies \cite{Ashraf_ES:2024,Tong:vldb_2018}, we assume that the road network is represented in the form of a grid $g$ that is divided into $n$ non-overlapping grid cells. 
The distance subgraph quantifies the distance between different grid cells. 
The vertices $V^D$ of this subgraph correspond to the grid cells, and the edges $E^D$ represent connections between adjacent grid cells. Each edge is assigned a weight $w^D_{(i)(j)}$, which signifies the distance between the central points of the connected grid cells.
Figure ~\ref{img:road} depicts a segment of the road network represented in terms of the distance subgraph. In this graph, the edge between vertices $v_3$ and $v_4$ has a value of $2$ which denotes that the distance between the central points of $v_3$ and $v_4$ is $2$ units. 

\textit{Customer-order subgraph $G^S$.}
This subgraph stores the predicted number of customer orders that are expected to arrive in food-delivery platforms. These predictions are generated using a graph neural network-based architecture, as described in \cite{Bhat:Cods_25}. 
The vertices $V^S$ of this subgraph represent grid cells, 
edges $E^S$ indicate the direction of the customer orders; specifically, if an edge exists between vertices $v_i$ and $v_j$, it implies that the customer is expected to place an order from restaurant $v_i$ and this order has to be delivered to $v_j$. Each edge is associated with a weight $w^S_{(i)(j)}$, which represents the predicted number of customer orders between the respective vertices.
Fig.~\ref{img:req} illustrates an instance of the customer-order subgraph. 
The edge weight between vertices $v_5$ and $v_4$ represented as $w_{(5)(4)}^S$ is $3$, 
which indicates that $3$ orders are expected to arrive at restaurant $v_5$ which have to be delivered to location $v_4$. \looseness=-1

\subsubsection{Problem formulation}


The primary objective of the proposed route recommendation system  is to create an eco-friendly platform wherein the delivery persons are placed in areas where the demand for food orders is expected to be high. 
The current approaches utilize a reactive approach \cite{Zhang:Springer_2022, Chen:IEEEITS_2024}, wherein delivery persons either wait passively or move around randomly in the search of orders. The proposed model overcomes this issue and  provides demand-aware route recommendation that aligns delivery person positions with future service needs. This method reduces idle cruising, improves the spatial efficiency of delivery person deployment, and lowers overall vehicular emissions.
To operationalize this, the proposed model utilizes a \textit{customer-order subgraph} $G^S = (V^S, E^S, w^S)$, where each node represents a spatial grid cell, and each directed edge $e^S_{(i)(j)} \in E^S$ carries a weight $w^S_{(i)(j)}$ that indicates the expected number of customer orders from location $i$ (e.g., a restaurant) to location $j$ (e.g., a customer). The goal is to identify a path within this subgraph that traverses areas with the highest anticipated order density, which will maximize the likelihood of a delivery person getting matched with an order shortly after repositioning.

However, repositioning over a long distance 
leads to increased fuel consumption and emissions that outweigh the environmental benefits of servicing more orders. To mitigate this, the system imposes a \textit{distance constraint} through \textit{distance subgraph} $G^D = (V^D, E^D, w^D)$, where the edge weights $w^D_{(i)(j)}$ capture the physical distances between neighboring vertices. Through this approach, the proposed model ensures that the recommended path remains within a predefined threshold while satisfying a higher count of customer orders. 
The optimization problem is formally defined as follows:


\begin{equation}
    P_{G^S}^{*}= \arg\max_{P_{G^S}} \mathbb{E}[|P_{G^S}|]
    \label{eq:obj}
\end{equation}

subject to
\begin{equation}
|P^{*}_{G^D}| <d_m
  \label{eq:distance_constraint}
\end{equation}

where
\begin{equation}
  |P_{G^S}| = \sum_{v_i \in P^S} \sum_{v_j \in \mathcal{F}} w^{S}_{(i)(j)}
  \label{eq:objsimplified}
\end{equation}
  \label{eq:length_road}

\begin{equation}
  |P^{*}_{G^D}| =    \sum_{\text{$(v_i, v_{i+1})$ in $P^{*}_{G^D}$}} w_{(i)(i+1)}
\label{eq:length_road}
\end{equation}

Eq. \eqref{eq:obj} represents the objective function of the proposed route recommendation system, which identifies the path $P^{*}_{G^S}$ from the customer-order subgraph $G^S$ that has the highest number of predicted orders.
The predicted count of orders along a path is computed as the sum of edge weights $w^{S}_{(i)(j)}$ from all the vertices $v_i$ in the path $P^S$ ($v_i \in P^S$) to their forward nodes ($v_j \in \mathcal{F}$), as defined in Eq. \eqref{eq:objsimplified}. For example, consider the path $\{v_5, v_3, v_4\}$ in Figure \ref{img:req}. The predicted count of orders in this path is equal to $w^S_{(5)(3)} + w^S_{(5)(4)} + w^S_{(3)(4)} = 1 + 3 + 2 = 6$.

To ensure operational feasibility, the model incorporates a distance constraint, as specified in Eq. \eqref{eq:distance_constraint}. This constraint ensures that the total distance of the recommended path, which is calculated through Eq. \eqref{eq:length_road} 
(with the distance subgraph $G^D$) 
does not exceed a predefined threshold $d_{\text{m}}$. 
This formulation encapsulates the primary objective of maximizing orders assigned to delivery persons, subject to a constraint which ensures that travel distance for repositioning remains within a bounded threshold. \looseness=-1

\subsubsection{Routing solution}

The demand for food delivery follows specific spatio-temporal patterns, and the delivery persons operating in these platforms are unaware of these patterns, which results in their placement in areas where demand is not high.  The proposed routing approach recommends the routes to delivery persons where anticipated food order demand is high 
which reduces the vehicle cruising in search of orders, 
and creates a sustainable system.

\begin{figure}[t!]
\vspace*{-9mm}
\centering
\begin{minipage}[b]{0.53\linewidth}
    \centering
    \includegraphics[width=0.9\linewidth]{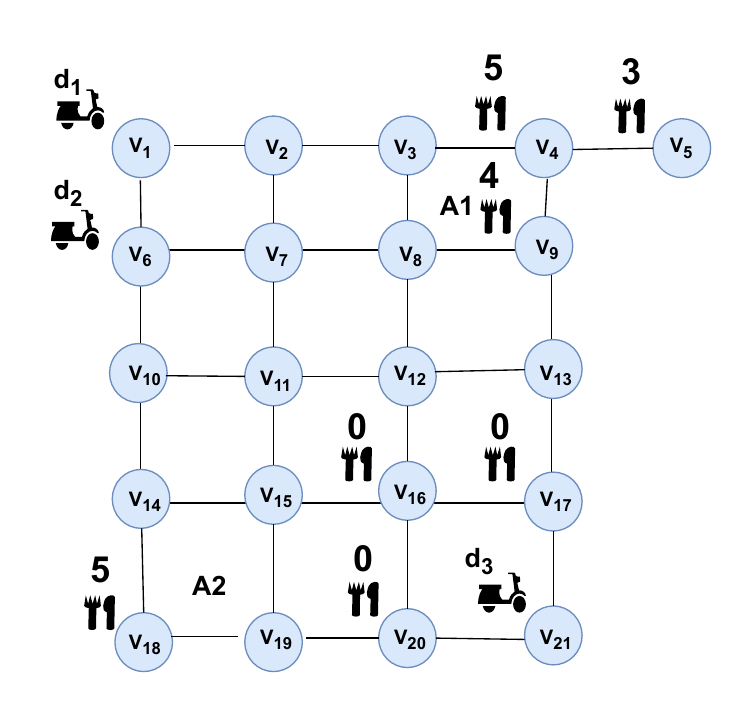}
    \caption{Road with vertices as grid cells and edges as connections}
    \label{img:ex1}
\end{minipage}
\hfill
\begin{minipage}[b]{0.45\linewidth}
    \centering
    \includegraphics[width=0.74\linewidth]{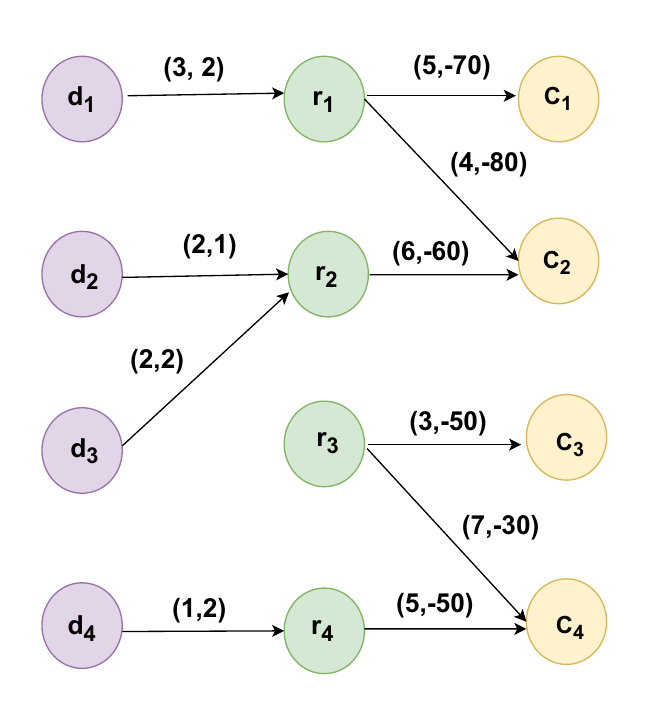}
    \caption{Three-layered order allocation graph}
    \label{img:flow_nw}
\end{minipage}
\vspace*{-7mm}
\end{figure}

To determine the recommended routes, the model selects paths with the maximum summation of edge weights in the customer-order subgraph as determined through Eq. \eqref{eq:obj}, where the edge weight represents the expected number of customer-orders between vertices. The path with the highest edge weight corresponds to the highest orders, which increases the vehicle utilization and optimizes the earnings of delivery persons. 
To illustrate this concept, consider Figure \ref{img:ex1}, where vertices represent grid cells which can include customer and restaurant locations, and edges represent the road segments.  
For the sake of illustration, assume there are three delivery persons initially positioned at vertices $v_1$, $v_6$, and $v_{21}$, and seven restaurants located at vertices $v_4$, $v_5$,$v_9$, $v_{16}$, $v_{17}$, $v_{18}$, and $v_{20}$. Among these restaurants, only the four located at $v_4$, $v_5$, $v_9$, and $v_{18}$ are active during the current time frame with $5$, $3$, $4$, and $5$ orders respectively.
From the figure, it is evident that delivery persons $d_1$ and $d_2$  are located in regions with sparse density of restaurants, while $d_3$  is situated in an area with dense restaurant coverage. However, despite the proximity of $d_3$ to numerous restaurants, the demand for food orders from these establishments is relatively low during the current time frame. Without a predictive mechanism, the delivery persons would remain unaware of alternative areas with higher anticipated demand, which would lead to inefficient cruising and underutilization of resources. To address this inefficiency, the proposed model first predicts the customer orders 
and thereafter recommends optimal routes to the delivery persons. 
In the context of Figure \ref{img:ex1}, the model predicts future demand (represented by vertex weights in the figure) 
and determines that relocating $d_1$, and $d_2$,  to area $A_1$, and  $d_3$ to area $A_2$, will maximize their likelihood of receiving orders. The recommended path ensures that delivery persons are routed through the shortest possible trajectories to these high-demand areas, which reduces vehicle movement and fuel consumption. \looseness=-1 

After analyzing the above example, we can see that the recommendation of efficient routes to delivery persons places them in areas where customer demand is expected to be high, which improves resource allocation and creates a sustainable system. 
However, selecting the paths with the highest orders on a directed graph is NP-Hard due to its reduction from the Longest Path Problem \cite{Ashraf_ES:2024}. To address this complexity, the proposed approach utilizes the structural properties of the objective function, and based on them, it applies the greedy algorithm for recommending routes.
The objective function, represented through Eq. \eqref{eq:obj}, is defined as a \emph{linear} combination of edge weights and exhibits two key properties: submodularity and monotonicity. 

\textit{Submodularity} captures the notion of diminishing returns, which implies that the marginal gain of adding an edge to a path decreases as more edges are included. Formally, a set function $f: 2^V \to \mathbb{R}$ is submodular if for any subsets $A \subseteq B \subseteq V$ and any element $e \in V \setminus B$, the following inequality holds: \looseness=-1
\[
f(A \cup \{e\}) - f(A) \geq f(B \cup \{e\}) - f(B)
\]
In our context, submodularity ensures that as more edges are added to a path, the incremental benefit of including additional edges diminishes. This property is particularly advantageous because it allows us to approximate the solution efficiently without exhaustively searching through all possible paths, which would be computationally infeasible for large graphs. Additionally, the submodular nature of the objective function aligns well with the practical constraints of real-world food delivery systems. For instance, delivery persons cannot traverse arbitrarily long distances to reach high-demand areas, as this would increase fuel consumption and reduce service efficiency. Submodularity inherently accounts for such limitations by ensuring that the inclusion of additional edges provides diminishing returns, thereby discouraging longer routes. This results in a sustainable and efficient system that minimizes vehicle movement while maximizing the likelihood of order fulfillment. \looseness=-1 

\textit{Monotonicity}, on the other hand, ensures that the value of the objective function does not decrease when new edges are added to the path. Formally, a function $f$ is monotone if for any subsets $A \subseteq B \subseteq V$, we have:
\[
f(A) \leq f(B)
\]
Our proposed objective function represented through Eq. \eqref{eq:obj} is monotone. This is because the customer-order values are always non-negative, and the addition of a non-negative value to a function does not decrease its value.
In the context of route recommendation, monotonicity ensures that the addition of more edges to a path does not reduce the total utility, which aligns with our goal of maximizing the total customer orders covered by the recommended path. 


The submodular and monotone nature of the objective function enables the efficient application of a greedy algorithm to the route recommendation problem. These structural properties ensure that locally optimal decisions yield strong approximation guarantees for the global optimum. At each step, the algorithm evaluates the immediate neighborhood of the current vertex and selects the adjacent edge with the highest predicted order count, as long as the cumulative distance constraint is not violated. This edge selection is performed over the customer-order subgraph, where edge weights represent the predicted demand. Algorithm \ref{alg:greedy_route_recommendation} (see Appendix \ref{sec:appendix1}) describes the working of greedy algorithm in detail. 

\vspace{-3mm}

\subsection{Order-allocation using Flow-based framework}

After moving the delivery persons to high demand areas, the proposed model allocates the orders to them through the network flow based approach.
Order allocation in food delivery platforms involves assigning available delivery persons to customer orders while minimizing operational costs, ensuring timely delivery, and promoting sustainability. Conventional approaches employ heuristic or rule-based systems that prioritize proximity between delivery persons and restaurants without considering the global structure of the system.
For instance, some of the models \cite{Tong:vldb_2018,Joshi:ACMTrans_2022} have minimized the travel distance between delivery persons and restaurants without balancing the delivery person workload or reducing vehicle movement. These limitations result in suboptimal resource utilization, where delivery persons are underutilized in low-demand areas while high-demand regions remain underserved. Furthermore, many existing approaches do not explicitly incorporate eco-friendly principles, such as reducing greenhouse gas emissions by minimizing vehicle cruising distances or relocating delivery persons to high-demand areas.
To address these limitations, we propose a novel \emph{Min-Cost Max-Flow (MCMF)} approach for order allocation in food delivery platforms. This framework builds upon the prediction and route recommendation mechanism described in the previous subsection, which positions the delivery persons in areas where the demand for food orders is expected to be high and uses this data to allocate the customer orders to delivery persons.  
The following subsection will discuss the graphical modelling and the MCMF approach for order allocation.  \looseness=-1

\textit{Graph-based modelling.}
The order allocation problem is represented as a three-layer directed graph:

\begin{equation*}
\vspace*{-1mm}
    G^O = \Big( \{V^{(i)}\}_{i \,\in\,\{1,2,3\}}, \{E^{(i,i+1)}, c^{(i,i+1)}, p^{(i,i+1)}\}_{i \,\in\,\{1,2\}} \Big)
\end{equation*}

where \( V^{(i)} \) is the set of vertices in layer \( i \), and \( E^{(i,i+1)} \subseteq V^{(i)} \times V^{(i+1)} \) represents the directed edges connecting consecutive layers. Each edge has an associated traversal cost

\begin{equation*}
    \vspace*{-1mm}c^{(i,i+1)}: E^{(i,i+1)} \to \mathbb{R}^{+}, \quad \forall i \in \{1,2\}
\end{equation*}

and a capacity constraint
\begin{equation*}
    p^{(i,i+1)}: E^{(i,i+1)} \to \mathbb{R}^{+}, \quad \forall i \in \{1,2\}
\end{equation*}

Next, we will provide a detailed description of the vertices and edges that constitute this structure. 

\textit{Vertices.}
The three layer vertex structure $\{V^{(i)}\}_{i \,\in\,\{1,2,3\}}$ consists of: \looseness=-1

\textit{Delivery person layer $ V^{(1)} $.}
This layer consists of nodes representing available delivery persons in the system. Each node corresponds to an individual delivery person who can be assigned to one or more customer orders based on his capacity. 

\textit{Pickup layer $ V^{(2)} $.}
This layer consists of nodes representing pickup locations, such as restaurants or dispatch points, where customer orders originate. These nodes function as intermediate points for assigning orders to delivery persons. 

\textit{Drop-off layer $ V^{(3)} $.}
This layer consists of nodes representing drop-off locations, such as customer addresses, where orders need to be delivered. These nodes 
represent the final stage of the assignment process. 

Figure \ref{img:flow_nw} shows an instance of the order-allocation graph with the vertices are divided into three layers: Layer $1$ - the delivery person layer representing the  $4$ available delivery persons $\{d_1,d_2,d_3,d_4\}$, Layer $2$ the pick-up layer denoting the $4$ active restaurants $\{r_1,r_2,r_3,r_4\}$, and Layer $3$ the drop-off layer corresponding to the $4$ customer request areas $\{c_1,c_2,c_3,c_4\}$.


After describing the vertex structure of the graph represented through three layers, we will complete the graph by modelling the \textit{edges} between different layers. 
 The edges of the graph capture the feasible transitions between layers and are designed to enforce \emph{delivery person capacity} and \emph{customer demand} constraints. 
 They are described next.



\textit{Delivery person-to-Pickup Edges $ E^{(1,2)} $}.
These edges represent the feasible transitions between delivery person layer $ V^{(1)} $ and pick-up layer $ V^{(2)} $. A transition is considered feasible if the delivery person from layer 1 can reach the restaurant represented through layer 2 without reducing the quality of the food. To ensure this, the proposed approach matches the delivery persons with restaurant(s) that lie within a threshold distance from it. This constraint, apart from improving customer satisfaction, prevents delivery persons from travelling longer distances to pick up orders, which optimizes resource utilization and reduces the effect of greenhouse emissions. 

After defining feasible edges, i.e., the edges that connect delivery persons and restaurants without reducing the food quality, we will describe the edge weights, which are represented in terms of cost and capacity. 
The cost of an edge represents the cost of assigning the vertex in layer $i$ to the vertex in layer $i+1$, and capacity represents the flow of the network, i.e., the maximum flow that can be sent through the edge. 
For the delivery person to pickup layer, the cost associated with each edge represents the distance between the delivery person's current location and the restaurant, and the capacity of these edges is determined by the maximum number of orders a delivery person can take simultaneously. 
 Figure \ref{img:flow_nw} displays an instance of the order-allocation graph. In this graph, the edge between vertex $d_1$ and $r_1$ has a cost of $3$ which displays that delivery person $1$ is at a distance of $3$ units from restaurant $1$. Similarly, the capacity of the edge between $d_1$ and $r_1$ is $2$, which shows that delivery person $1$ takes a maximum of $2$ orders. 

\textit{Pickup-to-Drop-off Edges $ E^{(2,3)} $}
These edges represent the feasible transitions between pickup nodes $ V^{(2)} $ (restaurants) and drop-off nodes $ V^{(3)} $  (customers). A transition is considered feasible if the delivery from a restaurant to a customer can be completed within operational constraints, such as \emph{distance} and \emph{time} constraints. Distance constraints state that the orders are only placed from restaurants that are located within a feasible delivery radius. Platforms like Zomato and Swiggy enforce this by displaying only nearby restaurants to customers, which prevents impractical pairings that could lead to delays, cold meals, and inefficient resource use \cite{Article:swiggy_dist}. 
The proposed model satisfies distance constraints
by connecting the restaurants to customers only if the distance between them is bounded within a predefined threshold value.

The second constraint for an edge to be considered feasible is the time constraint, which mandates that all orders must be delivered within the limits specified by the Service Level Agreement (SLA). In the case of single-order deliveries—where one delivery person is assigned to a single order from a restaurant to a customer—this constraint is inherently satisfied by directly enforcing a distance-based threshold, ensuring that delivery occurs within the permitted time frame. However, when a restaurant handles multiple concurrent orders, satisfying SLA constraints becomes non-trivial, particularly when these orders are geographically dispersed. Arbitrarily batching such orders may lead to excessive delays for some customers, thereby violating SLA requirements. To address this, the proposed model introduces a \emph{
detour ratio} metric that quantitatively captures the additional travel incurred when an order is delivered as part of a batch, compared to its direct delivery route. This ratio serves as a criterion for batching spatially cohesive and directionally aligned orders, ensuring that delivery efficiency does not come at the expense of service quality. \looseness=-1

Formally, for a restaurant \( r \) serving a batch of \( l \) customer orders \( \{o_1, o_2, \dots, o_l\} \), the drop-off sequence is determined based on the shortest path ordering among the delivery locations. The detour ratio for an order \( o_i \), where \( i \geq 2 \), is computed as the ratio between the cumulative distance travelled from the restaurant through the intermediate drop-off points \( o_1 \) to \( o_{i-1} \) and the direct shortest-path distance from the restaurant to \( o_i \). That is, the detour ratio is given by
\[
d_r(o_i) = \frac{D(r, o_1) + \sum_{j=1}^{i-1} D(o_j, o_{j+1})}{D(r, o_i)}
\]
where \( D(a, b) \) denotes the shortest path distance between locations \( a \) and \( b \). For instance, in a batch consisting of orders \( o_1, o_2, o_3 \), the detour ratio for \( o_3 \) is computed as the total travel distance from the restaurant to \( o_1 \), from \( o_1 \) to \( o_2 \), and then from \( o_2 \) to \( o_3 \), divided by the direct distance from the restaurant to \( o_3 \). A detour ratio close to one indicates that the inclusion of preceding orders in the delivery route does not introduce significant deviation from the direct path, implying strong spatial alignment. The model permits batching only if the detour ratio for each order in the batch remains below a specified threshold, thereby ensuring that the inclusion of multiple orders does not lead to SLA violations. This formulation enables the platform to exploit the operational benefits of batching, such as improved delivery efficiency and reduced travel costs, while maintaining compliance with strict time constraints imposed by the SLA.

After the description of feasible edges, we move on to describing the cost and capacity associated with the edges. The cost of an edge represents the negative profit for delivering the order from the restaurant to the customer. 
 This formulation aligns with the Min-Cost-Max-Flow framework, where minimizing the total cost corresponds to maximizing the platform’s overall profit. 
The capacity is determined by the number of orders originating from the pickup location and destined for the drop-off location. This modelling ensures that the total flow from pickup nodes to drop-off nodes satisfies the demand at each destination, while ensuring the profit is maximized. To understand the cost and capacity of this layer consider Figure \ref{img:flow_nw}, where the cost of edge connecting $r_2$ with $c_2$ is $-60$ which denotes that the delivery fee for the order is $60$ units. Its capacity is $6$, which denotes there are $6$ orders from customer area $c_2$ to restaurant $r_2$. 

After modelling the nodes and the edges of the graph, we will describe the order allocation mechanism on this graph. The order allocation mechanism allocates the delivery persons represented by layer 1 to the restaurants represented through layer 2 for servicing the customer orders represented by layer 3 with the following constraints:  
1) delivery person capacity should be fully utilized, 2) order-allocation should be fuel and cost-efficient, and 
3) customer demand should be covered. 
To solve the order-allocation with the above constraints, the proposed approach uses the Min-Cost-Max-Flow approach on the three-layer architecture graph structure, which is described layer-wise next. 

\textit{Layer 1 to Layer 2: Maximizing delivery person utilization while minimizing cost.}
The first phase of the proposed architecture addresses the assignment of delivery persons from layer 1 to the restaurants represented through layer 2, subject to the delivery person capacity and distance constraints. We know that the delivery persons are connected to restaurants that lie within a specified threshold distance to ensure customer satisfaction. 
The edges in this bipartite graph represent feasible transitions that satisfy operational requirements and uphold customer satisfaction. Once the feasible edges are identified, the proposed model determines the allocation of flow along these edges. The flow corresponds to the assignment of delivery persons to restaurants, and the proposed approach aims to optimize this assignment to satisfy their \emph{ capacity} constraints.  Specifically, the objective is to maximize the total number of orders assigned to delivery persons, ensuring that their capacity is not exceeded. 
However, the delivery persons are located at varying distances from the orders, and the proposed model ensures that they travel the least to be assigned to restaurants. 
This is achieved by selecting the flow that minimizes the cost of order allocation, where the cost is proportional to the distance between delivery persons and restaurants. 
This will ensure the delivery persons located closest are assigned orders which aligns with the modelling of the food delivery platforms and reduces the extra distance travelled by the delivery persons. 
This optimization is achieved through the Minimum-Cost Maximum-Flow framework, which maximizes the flow—representing the total number of orders assigned to delivery persons—while respecting their capacity constraints. 
\textit{Layer 2 to Layer 3: Demand-aware flow and profit maximization.}
Once delivery persons are allocated to pickup locations, the next phase involves routing them to customer destinations in layer 3. 
This step introduces a demand-aware flow mechanism, where delivery persons are dispatched based on the actual demand at drop-off locations. Specifically, if there is a food delivery request at a particular destination, the model routes delivery persons accordingly; otherwise, no flow is sent. This ensures that resources are utilized on demand, which optimizes the efficiency of the system. 
However, among all the flows, the proposed model aims to select the flow that maximizes the system profit. To optimize profit, the proposed approach uses the cost of each edge, which is defined as the negative of the delivery fare. It selects the edges with the least cost, which corresponds to the maximization of system profit. This framework directs the delivery persons from the restaurants to the customers based on the actual demand and optimizes the system profit. 

Through the combination of three layer architecture, the proposed approach allocates the customer orders to the delivery persons based on their capacity, the demand at each place, and the profit of platform. This framework ensures that the order allocation is sustainable while ensuring the customer service and the system profit are optimized, making it well-suited for real-world food delivery platforms. Algorithm \ref{alg:mcmf} (see Appendix \ref{sec:appendix2}) describes the working of the flow model in detail. \looseness=-1

\vspace{-3mm}
\section{Experiments and Results}
\label{sec:exp}
This section provides a comprehensive experimental evaluation of the proposed model and defines its operational efficiency when deployed in different parameter settings.

\vspace{-3mm}

\subsection{Experiment}
Firstly, we will describe the experimental setup used for validating the performance of the proposed approach.

\subsubsection{Experimental settings}
The experiments were conducted using Python on a machine equipped with an Intel\textsuperscript{\textregistered} Core\texttrademark{} i9-12900 CPU operating at 2400~MHz and 32~GB of RAM.

\subsubsection{Dataset description}
We evaluate our proposed model on the Miutenan dataset, which captures real-world food delivery operations. The dataset contains detailed logs of delivery requests, including timestamps, restaurant and customer order locations, delivery durations, and order volumes. Table \ref{tab:dataset-summary} provides a description of the dataset. 
The dataset was collected for the month of October 2022 and was discretized into spatial grid cells of size $2$ km and temporal intervals of $15$ minutes. The choice of a $2$ km grid resolution is motivated by its ability to effectively represent the complexity of urban road networks, as demonstrated in prior research on route recommendation systems \cite{Tong:vldb_2018,Ashraf_ES:2024}. The dataset provides spatio-temporal data about customer orders, which serves as input to the graph neural network-based model described in \cite{Bhat:Cods_25}. The model predicts the customer orders between any pair of locations for the upcoming $15$-minute interval 
which forms the basis for route recommendation and order allocation within the system.


\subsubsection{Metrics}
The performance of the proposed model is evaluated on the following metrics:

\textbf{Vehicle count.} This metric represents the total number of vehicles actively engaged in serving customer requests during the evaluation period. A lower vehicle count for the same level of service indicates better resource utilization.

\textbf{Efficiency.} It is measured as the number of orders taken by the platform, and is used to quantify the profit of the platform. 

 \textbf{Profit.} Profit of a driver is defined as the difference between the total revenue generated from completed orders and the associated operational cost which includes travel distance. 

\textbf{Service time.}  It refers to the total time taken from order placement to final delivery. Lower service time implies faster delivery and better customer satisfaction. 

\subsubsection{Baselines}
We evaluate the performance of the proposed model by comparing it with the following baselines.

\textbf{Order Bundling (OB)} \cite{Ye:ElsTR_2024}. It bundles multiple orders and enables courier sharing among restaurants.

\textbf{Branch and Bound (BB)} \cite{AGNETIS:ElsCI_2023}. It proposes a single courier meal delivery problem that uses branch and bound algorithm.

\textbf{Greedy.} It greedily assigns orders to the nearest available driver.
\looseness=-1


\vspace{-2mm}

\subsection{Results and Discussion}
This subsection first describes the working of the proposed model in different parameter settings and thereafter evaluates its performance against the existing baselines.

\begin{figure*}[t]
  \vspace*{-27mm}
    \centering
    \begin{subfigure}[b]{0.24\textwidth}
        \includegraphics[width=\linewidth]{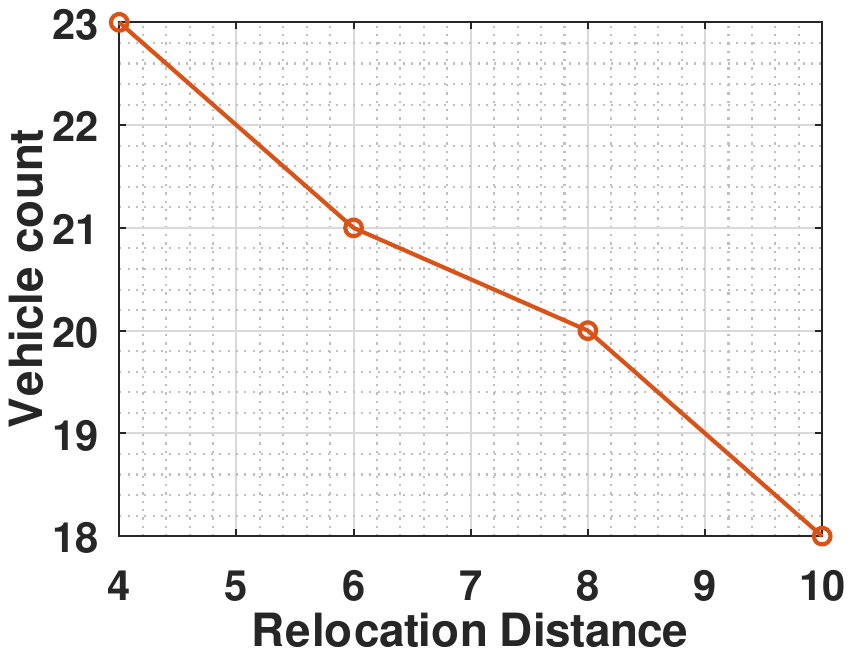}
        \vspace*{-22mm}
        \caption{Vehicle count}
        \label{img:vc}
    \end{subfigure}\hfill
    \begin{subfigure}[b]{0.24\textwidth}
        \includegraphics[width=\linewidth]{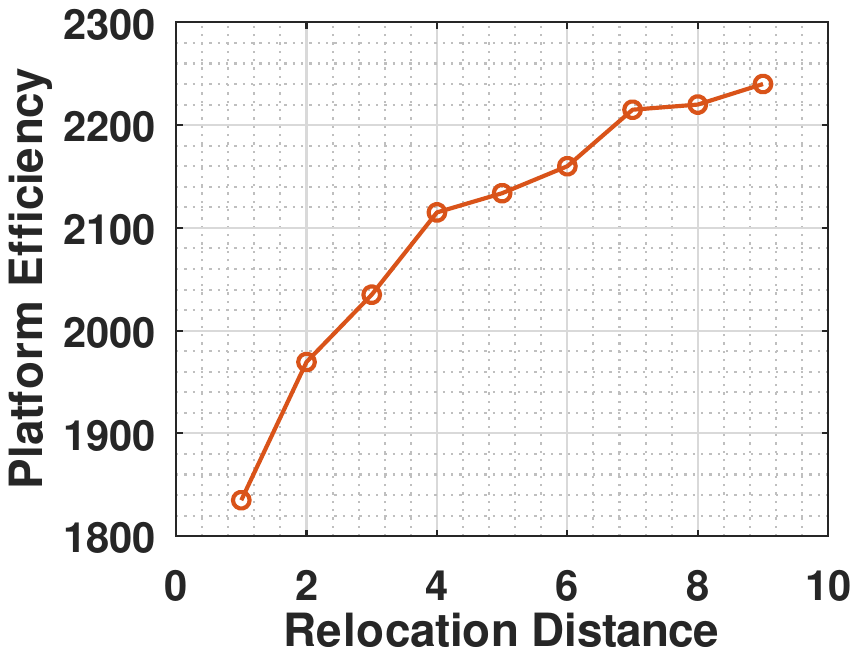}
      \vspace*{-22mm}  \caption{Efficiency}
        \label{img:eff_d}
    \end{subfigure}\hfill
    \begin{subfigure}[b]{0.24\textwidth}
        \includegraphics[width=\linewidth]{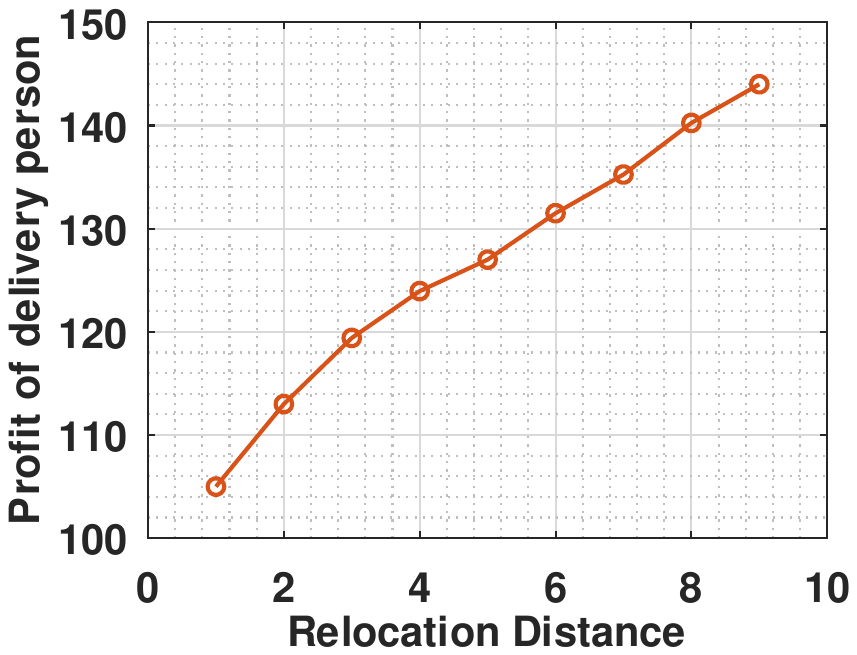}
        \vspace*{-22mm}\caption{Profit}
        \label{img:profit_d}
    \end{subfigure}\hfill
    \begin{subfigure}[b]{0.24\textwidth}
        \includegraphics[width=\linewidth]{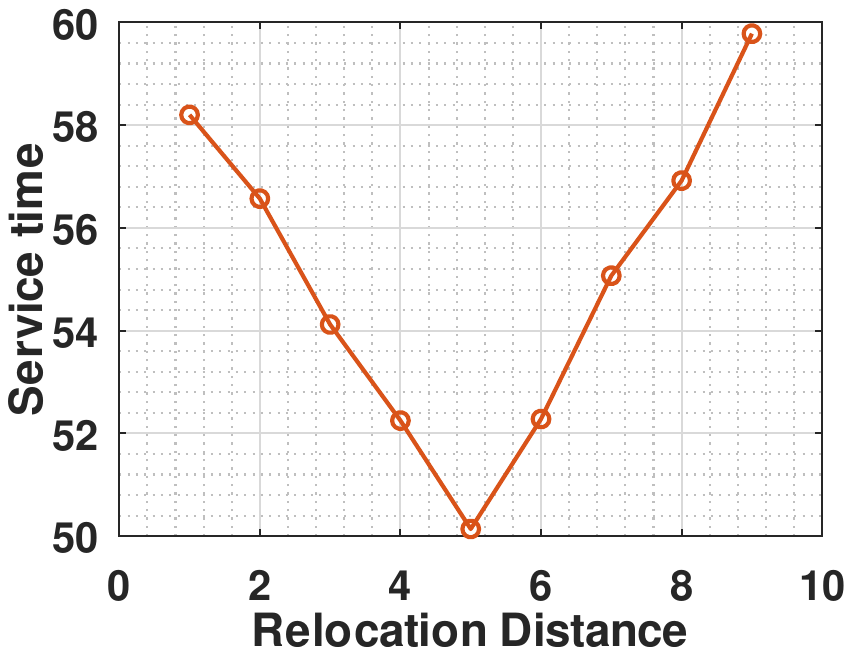}
        \vspace*{-22mm}\caption{Service time}
        \label{img:st_d}
    \end{subfigure}
    \vspace*{-1.5mm}
    \caption{Performance metrics evaluated on relocation distance}
    \label{utility}
\end{figure*}

\begin{figure*}[t]
  \vspace*{-22mm}
    \centering
    \begin{subfigure}[b]{0.24\textwidth}
        \includegraphics[width=\linewidth]{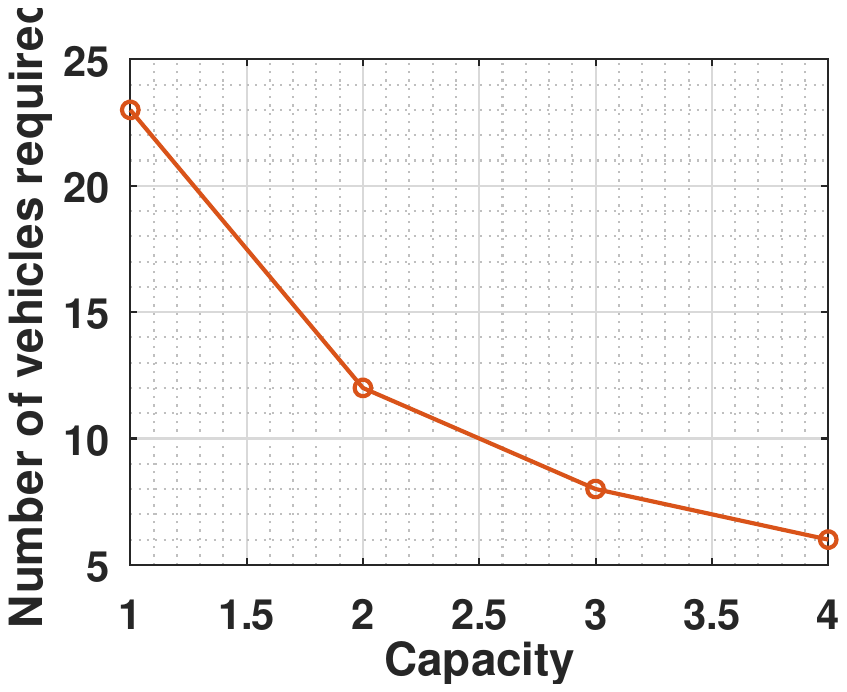}
         \vspace*{-22mm} \caption{Vehicle count}
        \label{img:vc_c}
    \end{subfigure}\hfill
    \begin{subfigure}[b]{0.24\textwidth}
        \includegraphics[width=\linewidth]{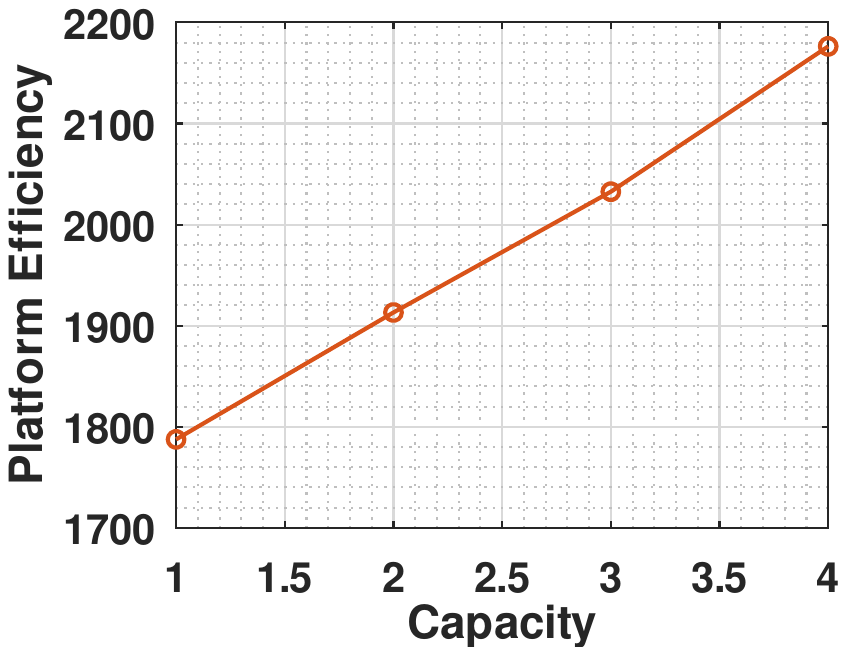}
       \vspace*{-22mm}   \caption{Efficiency}
        \label{img:eff_c}
    \end{subfigure}\hfill
    \begin{subfigure}[b]{0.24\textwidth}
        \includegraphics[width=\linewidth]{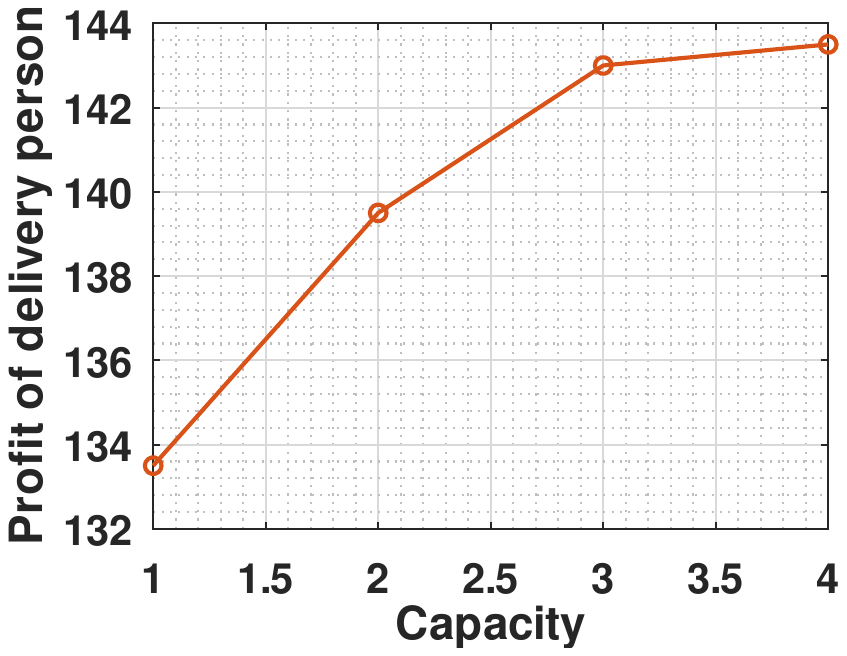}
        \vspace*{-22mm}  \caption{Profit}
        \label{img:profit_c}
    \end{subfigure}\hfill
    \begin{subfigure}[b]{0.24\textwidth}
        \includegraphics[width=\linewidth]{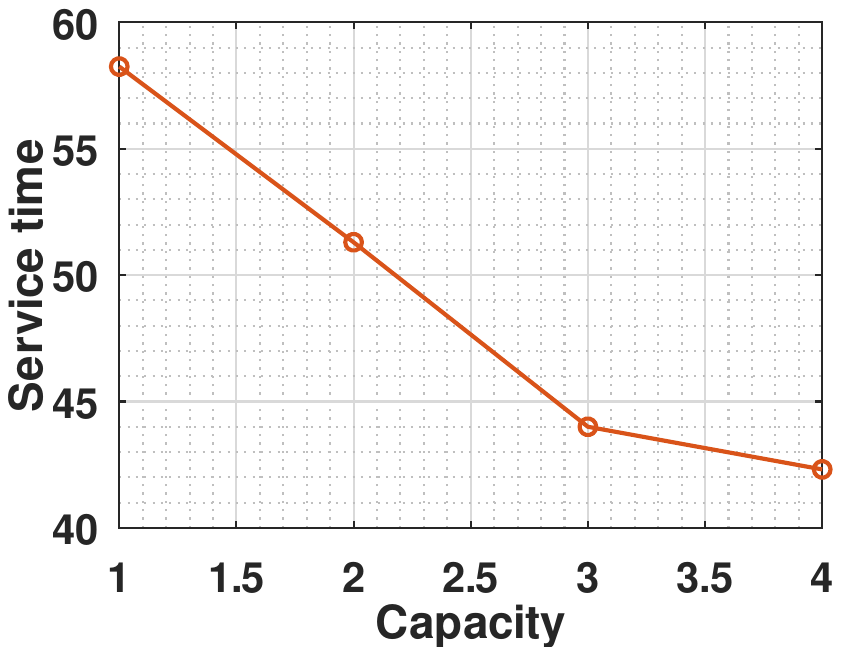}
         \vspace*{-22mm} \caption{Service time}
        \label{img:st_c}
    \end{subfigure}
   \vspace*{-1.5mm}  \caption{Performance metrics evaluated on capacity}
    \label{utility}
\end{figure*}

\begin{figure*}[t]
\vspace*{-22mm}
    \centering
    \begin{subfigure}[b]{0.24\textwidth}
        \includegraphics[width=\linewidth]{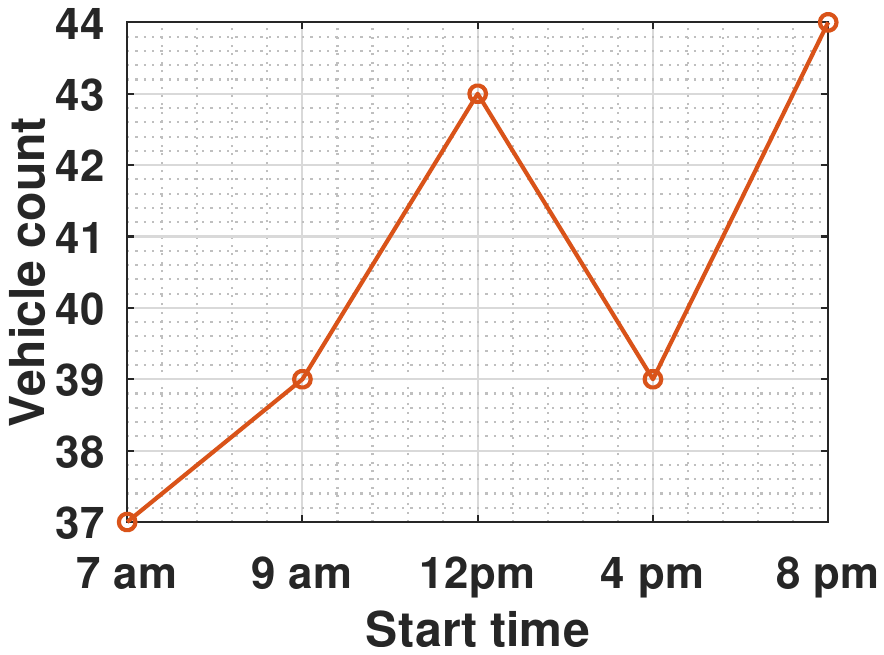}
     \vspace*{-22mm}    \caption{Vehicle count}
        \label{img:vc_t}
    \end{subfigure}\hfill
    \begin{subfigure}[b]{0.24\textwidth}
        \includegraphics[width=\linewidth]{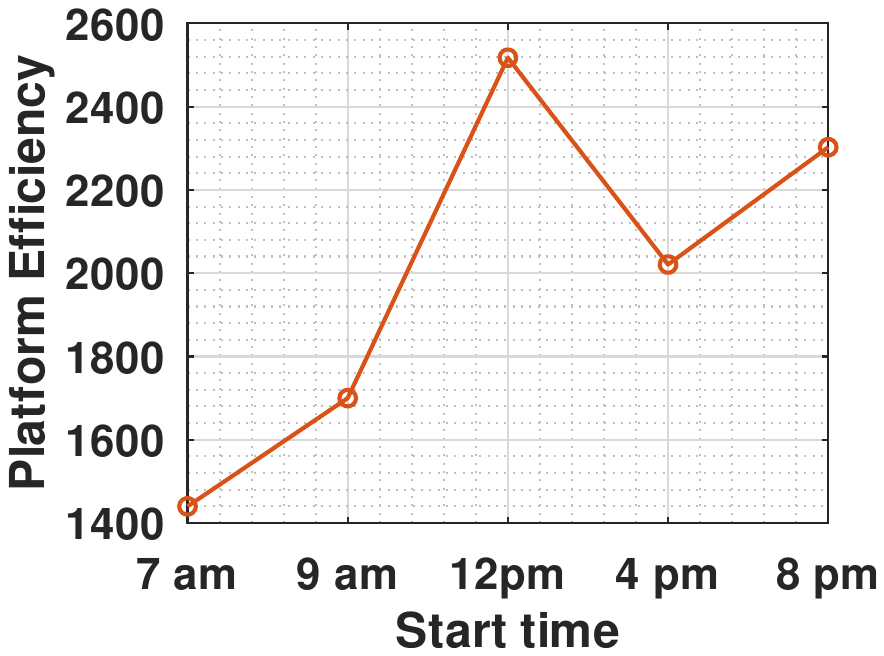}
      \vspace*{-22mm}   \caption{Efficiency}
        \label{img:eff_t}
    \end{subfigure}\hfill
    \begin{subfigure}[b]{0.24\textwidth}
        \includegraphics[width=\linewidth]{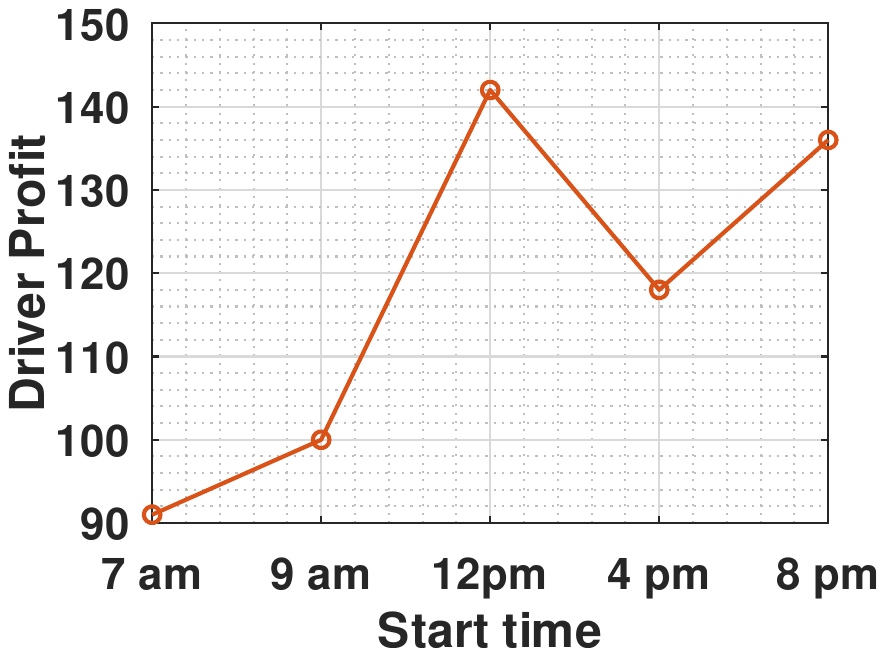}
       \vspace*{-22mm}  \caption{Profit}
        \label{img:profit_t}
    \end{subfigure}\hfill
    \begin{subfigure}[b]{0.24\textwidth}
        \includegraphics[width=\linewidth]{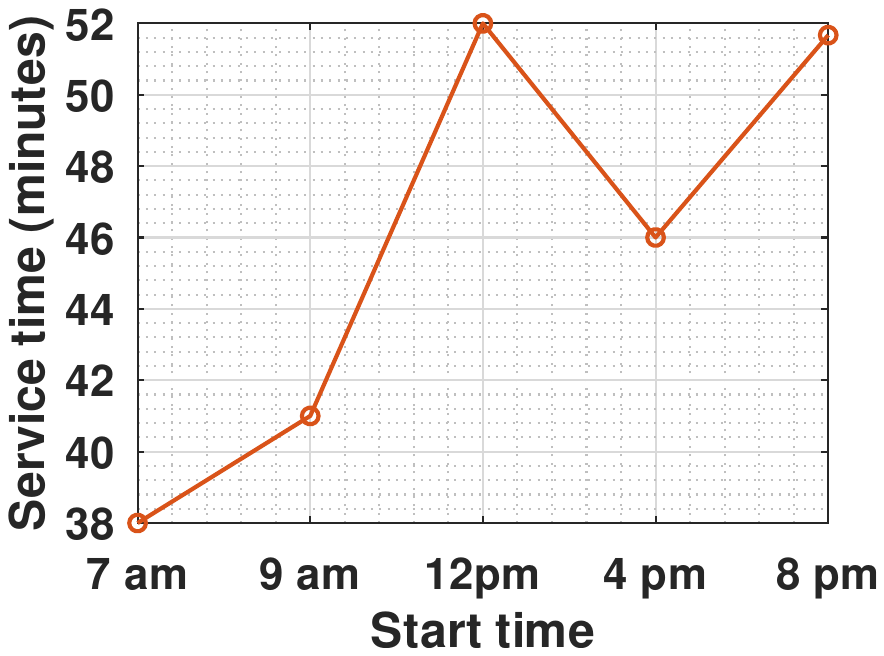}
    \vspace*{-22mm}     \caption{Service time}
        \label{img:st_t}
    \end{subfigure}
 \vspace*{-1.5mm}    \caption{Temporal evaluation of the proposed model}
    \label{utility}
\end{figure*}
\begin{figure*}[t]
 \vspace*{-22mm}
   \centering
    \begin{subfigure}[b]{0.24\textwidth}
        \includegraphics[width=\linewidth]{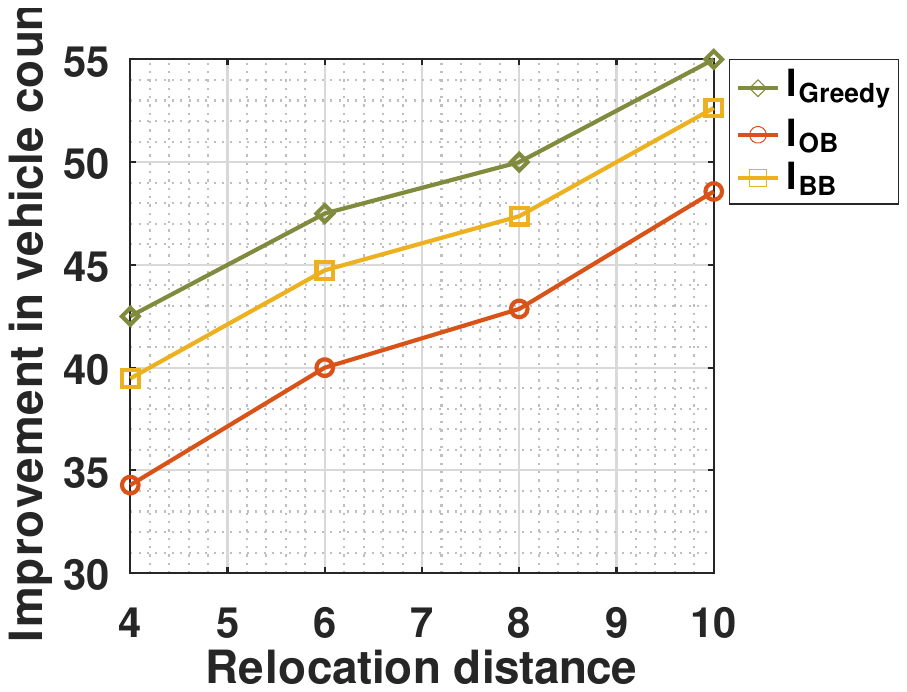}
     \vspace*{-22mm}   \caption{Vehicle count}
        \label{img:vc_b}
    \end{subfigure}\hfill
    \begin{subfigure}[b]{0.24\textwidth}
        \includegraphics[width=\linewidth]{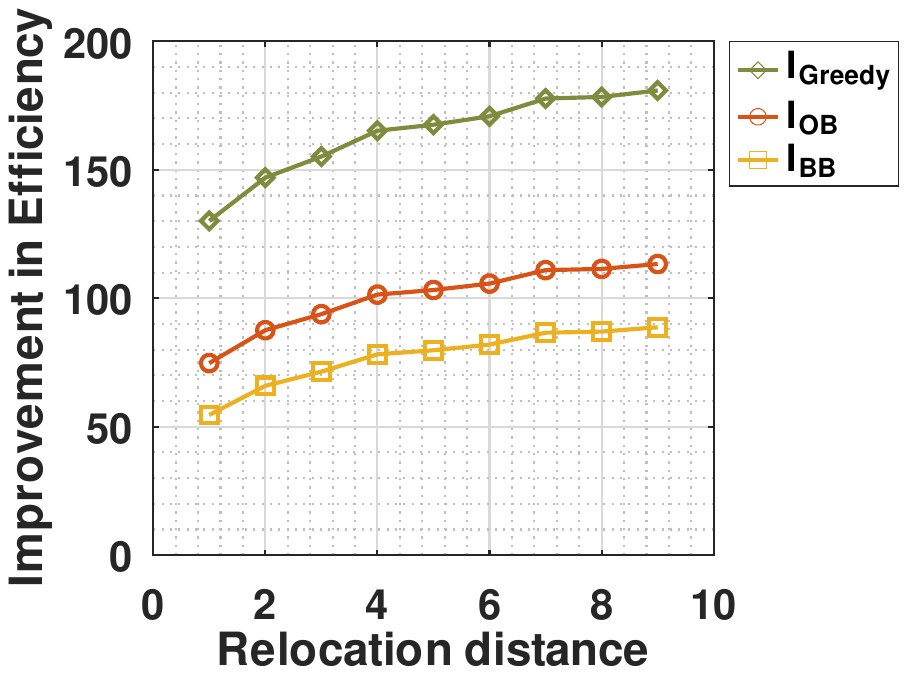}
     \vspace*{-22mm}    \caption{Efficiency}
        \label{img:eff_b}
    \end{subfigure}\hfill
    \begin{subfigure}[b]{0.24\textwidth}
        \includegraphics[width=\linewidth]{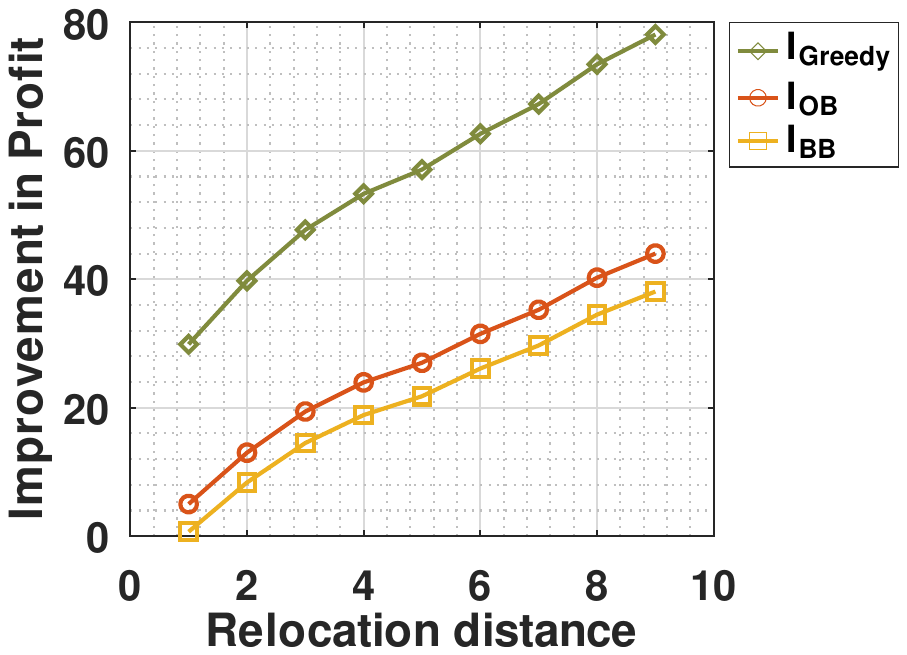}
       \vspace*{-22mm}  \caption{Profit}
        \label{img:profit_b}
    \end{subfigure}\hfill
    \begin{subfigure}[b]{0.24\textwidth}
        \includegraphics[width=\linewidth]{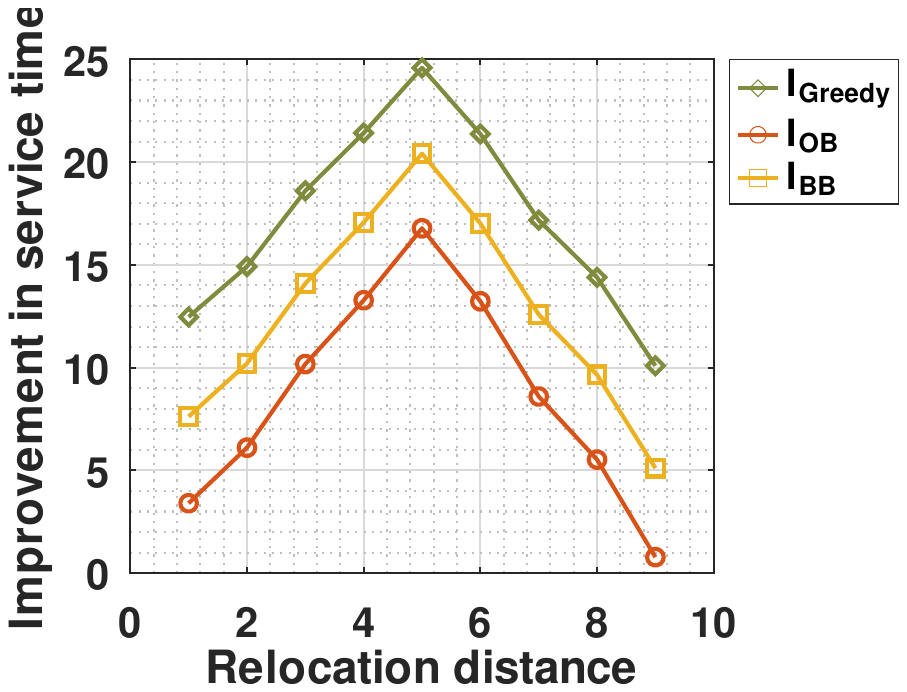}
      \vspace*{-22mm}   \caption{Service time}
        \label{img:st_b}
    \end{subfigure}
    \vspace*{-1.5mm} \caption{Comparison with the existing baselines}
    \label{utility}
    \vspace*{-6mm}
\end{figure*}

\subsubsection{Parameters}
There are two parameters for the proposed model- \emph{relocation distance} and the \emph{delivery person capacity}.

\textit{Relocation distance} quantifies the maximum distance within which the delivery persons are moved from their current position in the anticipation of getting a higher number of customer orders. 
    Figure \ref{img:vc} shows the effect of relocation distance on the vehicle count. With the increase in the relocation distance, the number of vehicles required to service the same volume of orders decreases significantly. This is because a larger distance aligns the spatial redistribution of agents with the predicted demand, which optimizes fleet utilization. When relocation is constrained to a small radius, delivery agents remain congested in low-demand areas, which shows up in a larger fleet size. 

 Figure \ref{img:eff_d} highlights the improvement in platform efficiency with the increase in relocation distance. 
 The higher relocation distance leads to better spatial matching between agents and customer orders as delivery agents begin their shifts closer to expected order origins, which reduces average travel time per order and displays in increased platform efficiency. 
Similar to efficiency, the profit of the delivery persons also increases with the increase in relocation distance, as can be seen through Figure \ref{img:profit_d}. 
When delivery agents are repositioned to high-demand zones, they experience reduced idle time and the likelihood of receiving bundled orders increases. This results in the minimization of downtime between successive deliveries and raises the per-hour profitability for drivers, which contributes to a more equitable and efficient labor model.

Figure \ref{img:st_d} demonstrates that a non-monotonic relationship exists between relocation distance and average service time. While the initial increase in relocation distance reduces service delay by improving demand-supply matching, excessive relocation beyond the optimal threshold introduces overhead travel time, partially offsetting the gains. This is because the increased relocation of drivers to high-demand areas causes accessibility issues in lower-demand areas, which results in an increased average service time. Though a higher relocation distance increases the profit of the driver and platform but it becomes unfair for the customers. To overcome this issue and create a balance between the two cases, we have set the default value of relocation distance as $5$ \textit{km}. 

\textit{Delivery person capacity.}
 Standard food delivery configurations often limit delivery persons to a single order to avoid complexity and SLA violations. However, this leads to an inflated number of trips and destabilizes the environment. Increasing the capacity to carry multiple compatible orders can improve vehicle utilization, reduce emissions, and increase delivery person profit, provided SLA and detour constraints are respected. This subsection describes the effect of vehicle capacity on different performance metrics.
Figure \ref{img:vc_c} shows the relation between capacity and the vehicle count. As can be seen through the figure, increasing delivery person capacity substantially reduces the number of vehicles required. This is because higher capacity facilitates multi-order bundling, which enables a single agent to serve several customers in a single trip. This compresses the delivery graph's operational footprint and allows the platform to service higher order volumes with fewer active vehicles.

 Figures \ref{img:eff_c} and \ref{img:profit_c} show the improvement in platform efficiency and driver profit with the increase in capacity. This is because a higher capacity places more orders per vehicle, which increases the order allocation of delivery persons and shows up in higher profit and system efficiency.

In a similar manner, service time (see Figure \ref{img:st_c}) decreases with an increase in capacity due to the higher availability of vehicles. This is because with an increased capacity, each vehicle can service multiple customers, which shows up in higher available vehicle count and reduces average service time. \looseness=-1


\subsubsection{Temporal analysis of platform performance}

To evaluate the robustness of the proposed framework under varying temporal demand patterns, we analyze its behavior across multiple time intervals throughout the day. This temporal decomposition is critical in food delivery platforms, where order density and geographic spread fluctuate significantly based on customer routines, restaurant schedules, and urban traffic dynamics.
In particular, we evaluate the performance of the proposed model  during peak demand periods—namely lunch (12 pm) and dinner (8 pm)—as well as during off-peak hours (7 am, 9 am, and 4 pm).


Figure \ref{img:vc_t} shows the vehicle count required by the proposed model at different time frames. The number of active delivery vehicles exhibits clear temporal variability. During off-peak hours (7 am, and 9 am), vehicle count remains relatively low, as the order volume is minimal and demand is spatially sparse. However, a marked increase is observed during core meal periods—specifically around 12 pm (lunch) and 8 pm (dinner). The rise in vehicle count during these intervals is a function of high order volumes combined with limited bundling feasibility due to temporal urgency and diverse delivery locations. Despite this, the proposed framework successfully restricts the peak vehicle requirement to a moderate level, indicating its ability to maintain high service level without oversaturating the fleet. \looseness=-1

Efficiency and profit displayed through Figures \ref{img:eff_t} and \ref{img:profit_t},
vary significantly across time frames. The platform operates at suboptimal efficiency in the early morning (7 am, and 9 am) due to order sparsity and long inter-pickup distances. Efficiency improves significantly during the 12 pm and 8 pm window, where increased demand density enables spatial bundling and route consolidation. 
The efficiency observed around 4 pm is particularly insightful—it indicates a transitional window where demand is substantial, but not too high, which allows the platform to aggregate orders with minimal detour. \looseness=-1


Figure \ref{img:st_t} shows the service time at different time frames. The service time is minimal during early hours ( 7 am, and 9 am), due to lower quantity of orders, less road congestion and minimal order overlap. However, the sharp rise around 12 pm and 8 pm is a consequence of high demand saturation. Despite increased bundling during these intervals, the rise in concurrent requests strains system responsiveness, leading to marginally higher delivery delays. The system still contains the average service time within acceptable operational threshold, due to its detour-aware batching and repositioning mechanism. \looseness=-1


\subsubsection{Comparative analysis with baseline approaches}

To assess the effectiveness of the proposed integrated framework, we compare its performance against three baseline approaches—Greedy, OB, and BB—across four performance metrics: (a) vehicle count, (b) efficiency, (c)  driver profit, and (d) service time. 
For each metric. we calculate the improvement over baselines which is defined as:

$\mathrm{I_A} =
\begin{cases}\tiny
    \left(\frac{ A-P}{P} \times 100\right) & \text{if metric denotes service time}  \\
     \left(\frac{ P-A}{P} \times 100\right) & \text{otherwise}
\end{cases}
$

Here, $\mathrm{I_A}$ represents the improvement of proposed model over baseline 
$A$, and 
$P$ represents the performance of the proposed model. The interpretation of each metric depends on its context: for efficiency, profit, and vehicle count, a higher value signifies better performance, while for metrics such as service time, a lower value displays better results.


Figure \ref{img:vc_b} displays the vehicle count required by the proposed model and the existing baselines. The proposed model achieves significant improvement in reducing the number of active delivery vehicles required to fulfil a given demand volume. The improvement increases consistently with relocation distance, which can be attributed to the model’s predictive repositioning mechanism, that aligns delivery persons with spatially distributed, high-demand regions in advance.
In contrast, the Greedy baseline employs myopic, locally optimal decisions, while OB and BB do not incorporate proactive repositioning based on customer demand. 
The proposed approach mitigates these inefficiencies through jointly optimized spatial realignment and order allocation, which leads to more compact and sustainable fleet usage.

Platform efficiency displayed through Figure \ref{img:eff_b}, shows substantial enhancement over the existing baselines. 
The observed performance gains stem from proposed model’s ability to simultaneously account for anticipated demand density and spatial alignment of customer orders. This indicates that the proposed model not only improves environmental efficiency but also creates a profitable system, which is important for the long-term sustainability in the market.
Similar to efficiency, delivery persons' profit (see Figure \ref{img:profit_b}), increases substantially under the proposed framework. 
This is primarily due to reduced idle time and increased task density per unit time, which shows up in higher values.

Apart from profit and efficiency, the proposed model improves the service time over the existing baselines. Figure \ref{img:st_b} shows average customer service time with an increase in relocation distance. The improvement initially increases with the relocation distance which can be attributed to the placement of delivery persons in the areas where demand is anticipated to be high. However, with the higher relocation distance the time to service an order increases due to the lower density of delivery persons operating in low-demand areas, which leads to an increased value of average time serviced.   

The experimental results demonstrate that the proposed model consistently outperforms baselines across all operational metrics. The improvements are a direct consequence of its joint optimization of delivery person repositioning and order allocation, which are based on the prediction system. 
Through this mechanism, the proposed model captures the spatial and temporal variations in order demand and shows up in improvement over the existing baseline approaches.


\vspace*{-2mm}

\section{Conclusion}
\label{sec:concl}
This paper introduced a unified, sustainability-driven framework for food delivery platforms that jointly optimized delivery person routing and order allocation. The proposed model integrates demand prediction, delivery person repositioning, and
multi-order allocation into a cohesive pipeline and establishes a proactive, demand-aware coordination strategy. It uses submodular optimization for repositioning drivers, and aligns them in areas projected with high demand. Thereafter, it utilizes a capacity and cost-aware min-cost max-flow allocation mechanism that enables the system to achieve high delivery efficiency with reduced environmental footprint.
The results demonstrate that aligning delivery logistics with anticipated demand patterns and spatial constraints optimizes vehicle usage, increases profitability, and improves service quality. This work highlights the importance of integrated, learning-based optimization in shaping the next generation of urban food delivery systems that are not only efficient, but also environmentally and operationally sustainable.

\bibliographystyle{IEEEtran}
\bibliography{sample-base}
\vspace{-6mm}
\appendices
\section{ Algorithm for Route Recommendation System}
\label{sec:appendix1}
\begin{algorithm}[H]
\caption{Greedy Route Recommendation}
\label{alg:greedy_route_recommendation}
\begin{algorithmic}[1]
    \Require Distance and customer-order subgraphs, threshold distance $d_{\text{m}}$
    \Ensure Recommended path $P_{G^S}^*$
    
    \State $P_{G^S}^* \gets []$ \Comment{Path to be constructed}
    \State $v_{\text{c}} \gets v_{\text{start}}$ \Comment{Starting node}
    \State $d_{\text{c}} \gets 0$ \Comment{Cumulative travel distance}
    
    \While{$d_{\text{c}} < d_{\text{m}}$}
        \State $\text{FeasibleEdges} \gets []$
        \For{each neighbor $v_j$ of $v_{\text{c}}$}
            \If{$d_{\text{c}} + w^D_{(v_{\text{c}})(v_j) }\leq d_{\text{m}}$}
                \State Add edge $(v_{\text{c}}, v_j)$ to $\text{FeasibleEdges}$
            \EndIf
        \EndFor
        \If{$\text{FeasibleEdges} = \emptyset$}
            \State \textbf{break} \Comment{No feasible extension}
        \EndIf
        \For{each edge $ (v_{\text{c}}, v_j)$ in $\text{FeasibleEdges}$}
           \State $\text{MarginalGain}[(v_c,v_j)] \gets w^S_{(v_{\text{c}}) (v_{j})}$

        \EndFor
        \State $(v_c,v_j^*) \gets \arg\max_{(v_c,v_j)} \text{MarginalGain}[e]$
\State Append $(v_c,v_j^*)$ to $P_{G^S}^*$
\State $v_{\text{c}} \gets v_j^*$ \Comment{Update current node}
\State $d_{\text{c}} \gets d_{\text{c}} + w^D_{(v_{\text{c}})( v_j^*)}$



    \EndWhile
    \State \Return $P_{G^S}^*$
\end{algorithmic}
\end{algorithm}
Algorithm 1 illustrates the working of the proposed greedy algorithm. 
 It operates over two subgraphs: the customer-order subgraph $G^S$, where edge weights represent the predicted number of orders between locations, and the distance subgraph $G^D$, which encodes the physical distance between adjacent regions. The core objective of the algorithm is to identify a path from the delivery person’s current location that maximizes the expected order count while ensuring that the total travel distance does not exceed a predefined threshold $d_m$.
The algorithm begins with an empty path and initializes the current node $v_c$ as the delivery person’s location. It iteratively explores neighboring nodes connected to the current location, and filters out those whose inclusion would cause the cumulative distance $d_c$ to exceed the threshold. For each feasible neighboring edge, it computes the marginal gain, which is defined as the predicted number of orders $w^S_{(v_c), (v_j)}$ between the current node $v_c$ and the neighboring node $v_j$. Among all feasible neighbors, the algorithm selects the edge with the highest marginal gain and extends the current path by appending this edge.
After each selection, the algorithm updates the current node to the newly selected vertex, increments the cumulative travel distance by the corresponding edge weight from $G^D$, and repeats the process. This greedy selection ensures that, at every step, the algorithm makes the most locally optimal choice in terms of predicted demand coverage. The loop terminates when no further feasible neighbors exist, either because the remaining edges exceed the distance constraint or all adjacent regions have been visited.
Through the construction of the incremental path and selection of the most rewarding edge at each step, the algorithm efficiently directs the delivery person through areas of high expected demand without the need for exhaustive search. This approach results in the creation of a greener system through the efficient placement of delivery persons at the location where the orders are expected to be high. 

\section{ Algorithm for Order Allocation}
\label{sec:appendix2}
\begin{algorithm}[H]
\caption{Min-Cost Max-Flow Algorithm}
\label{alg:mcmf}
\begin{algorithmic}[1]
    \Require Graph $G = (V^{(i)}, E^{(i,i+1)})$ with capacities $p^{(i,i+1)}$ and costs $c^{(i,i+1)}$ 
    \Ensure Maximum flow $f$ and minimum total cost
    
    \State Initialize:
    \State \quad Flow $f(u, v) \gets 0$ for all $(u, v) \in E$
    \State \quad Construct residual graph $G_f$ with residual capacities $r(u, v) \gets p(u, v)$
    
    \While{there exists an augmenting path $P$ from $s$ to $t$ in $G_f$}

\State Use Bellman-Ford to find the shortest path $P$ from $s$ to $t$ in $G_f$, minimizing $\sum_{(u, v) \in P} c(u, v)$
        
        \State Compute bottleneck capacity:
        \State \quad $\Delta \gets \min_{(u, v) \in P} r(u, v)$
        
        \State Augment flow along $P$:
        \For{each edge $(u, v) \in P$}
            \State $f(u, v) \gets f(u, v) + \Delta$
            \State $f(v, u) \gets f(v, u) - \Delta$
            \State Update residual capacities:
            \State \quad $r(u, v) \gets r(u, v) - \Delta$
            \State \quad $r(v, u) \gets r(v, u) + \Delta$
        \EndFor
    \EndWhile
    
    \State Compute total cost:
    \State \quad TotalCost $\gets \sum_{(u, v) \in E} f(u, v) \cdot c(u, v)$
    
    \State \Return Maximum flow $f$ and TotalCost
\end{algorithmic}
\end{algorithm}

Algorithm 2 describes the working of the proposed order allocation algorithm using the network flow architecture. 
It begins with an initialization step where the flow on all edges is set to zero, and a residual graph is constructed with capacities identical to the original network. The algorithm iterates while an augmenting path exists from the delivery person layer to the drop-off layer in the residual graph. At each iteration, the Bellman-Ford algorithm is employed to identify the shortest path in terms of cost. Once an augmenting path is determined, the algorithm computes the bottleneck capacity, which represents the minimum residual capacity along the path. The flow is then augmented by increasing it along the forward edges and decreasing it along the reverse edges to maintain the feasibility of future adjustments. The residual graph is updated accordingly by reducing the residual capacity on forward edges and increasing it on reverse edges. This process ensures that previous flow assignments can be modified if a more cost-effective path emerges in subsequent iterations.
The algorithm continues this iterative process until no further augmenting paths can be found, ensuring that the maximum possible flow is routed from the delivery person layer to the drop-off layer at minimal cost. Upon termination, the total cost is computed by summing the product of flow values and their corresponding edge costs across the network. The final output consists of the maximum flow and its associated minimum cost. This structured approach ensures an optimal allocation of flow while maintaining efficiency in cost minimization.

\section{Experimental data}
\begin{table}[h!]
\centering
\caption{Key characteristics of the Miutenan dataset}
\begin{tabular}{|l|p{5cm}|}
\hline
\textbf{Aspect} & \textbf{Details} \\
\hline
Month/Year &  October/2022 \\
\hline
Total Orders & ~ 569 million \\
\hline
Spatial Data & Pickup and drop-off locations  \\
\hline
Grid size & $2$ km \\
\hline
Time length & $15$ minutes \\
\hline
\end{tabular}
\label{tab:dataset-summary}
\end{table}

\end{document}